# Prediction of financial time series using LSTM and data denoising methods

Qi Tang[1], Tongmei Fan[2], Ruchen Shi[3] and Jingyan Huang[5]

School of Economics and Management, Southeast University

Nanjing, China

[1]regantangqi@seu.edu.cn

[2]m18851669637@163.com

[3]ruchenshi@gmail.com

[5]213192750@seu.edu.cn

Yidan Ma[4]

School of Cyber Science and Engineering, Southeast University

Nanjing, China

[4]ma726389@iCloud.com

Communicated by Wei Yin

In order to further overcome the difficulties of the existing models in dealing with the non-stationary and nonlinear characteristics of high-frequency financial time series data, especially its weak generalization ability, this paper proposes an ensemble method based on data denoising methods, including the wavelet transform (WT) and singular spectrum analysis (SSA), and long-term short-term memory neural network (LSTM) to build a data prediction model, The financial time series is decomposed and reconstructed by WT and SSA to denoise. Under the condition of denoising, the smooth sequence with effective information is reconstructed. The smoothing sequence is introduced into LSTM and the predicted value is obtained. With the Dow Jones industrial average index (DJIA) as the research object, the closing price of the DJIA every five minutes is divided into short-term (1 hour), medium-term (3 hours) and long-term (6 hours) respectively. . Based on root mean square error (RMSE), mean absolute error (MAE), mean absolute percentage error (MAPE) and absolute percentage error standard deviation (SDAPE), the experimental results show that in the short-term, medium-term and long-term, data denoising can greatly improve the accuracy and stability of the prediction, and can effectively improve the generalization ability of LSTM prediction model. As WT and SSA can extract useful information from the original sequence and avoid overfitting, the hybrid model can better grasp the sequence pattern of the closing price of the DJIA. And the WT-LSTM model is better than the benchmark LSTM model and SSA-LSTM model.

## 1. Introduction

The U.S., the largest economy in the world, has a perfect statistical system and the best financial supervision system, whose financial data features are complete, accurate and reliable. At the same time, the U.S. stock market and other U.S. capital markets cooperate in an efficient way and are an important part of the U.S. financial system, and all these characteristics make the market a good model. On the one hand, global stock markets react quickly and follow the trend of the U.S. market, especially in the case of unusually high market volatility. On the other hand, most economic theories and assumptions are based on the study of developed financial systems with larger and more liquid stock markets, more mature economies and more effective financial regulatory systems. As a representative of developed markets, American market is also the most favorable object of empirical or theoretical proposition in academic research. The three major stock indexes in the United States are the Dow Jones Index [1], the Standard & Poor's 500 Index and the Nasdaq Composite Index. The most famous of these indexes is the Dow Jones Index. The importance of the Dow Jones Index has been further recognized in global markets, beyond its role in the domestic market. The 30 companies that make up the Dow, such as Bank of America, Coca-Cola, General Electric, Microsoft and others, are well-known multinational corporations. They cover a wide range of large industries and their performance is behind the global economy. Therefore, forecasting the Dow Jones index is of great significance to the entire financial system.

At present, financial time series forecasting models can be divided into two categories: parametric model and non-parametric model. Autoregressive (AR), moving average (MA), autoregressive moving average (ARMA) and autoregressive comprehensive moving average (ARIMA) are typical models of parameter types [2]. All of the above models have some statistical assumptions, that is, they can only be used if the predicted time series satisfies the assumptions. Therefore, the parametric model has limitations to some extent. In addition, the parametric model is more suitable for the time series with linear characteristics, but the time series of DJIA price is full of non-linearity and high volatility which This will cause too many estimated parameters and increase the complexity of the model. Due to the limitations of the parametric model and characteristics of DJIA price, the parametric model is not suitable for Dow Jones index price forecast.

Stock price forecasting is a relevant and challenging problem that has attracted the interest of many engineers and scientists. In order to adapt to the characteristics of DJIA and achieve high precision, non-parametric models are used. Specifically, machine learning (ML) and deep learning (DL) models are used for DJIA price prediction. Ahmed [3] found that machine learning methods can perform better than traditional econometrics methods. Sert et al. [4] have proposed to combine various natural language processing and machine learning techniques and applied the developed methodology on stock market related case study. According to the results, they observed that if the features are selected logically, reasonable models that have more than 60% average accuracy can be created. This approach can be easily applied to create a prediction model from any corpus of textual data. Using these additional analysis techniques can enrich our feature set and improve the performance of the prediction models. Alfonso et al. [5] have proposed two methods to predict stock price and price intervals respectively. Both approaches are highly parallelizable, thus making possible to manage large data sets. Both techniques have been proved to be useful for the investors in terms of accuracy, and have other advantages like adaptation to the current market situation. Thus, the proposed techniques have proved that they can be added to the toolbox of stock market traders. In addition, the proposed approaches have been applied to the k-step forecasting of the Dow Jones Industrial Average index. However one would be the study of which technical indicators work best as market state, i.e., a feature selection study tailored for the proposed approaches. the suitability of these technical indicators with the proposed techniques is something that needs to be studied in future works. Chun et al. [6] have presented a technique called geometric Case Based Reasoning. This method overcomes the limitation of conventional case-based reasoning in that it uses Euclidean distance and does not consider how nearest neighbors are similar to the target case in terms of changes between previous and current features in a time series. The results show that the proposed technique is significantly better than the random walk model at $p < 0.01$. The proposed method has the possibility to improve predictability. At the same time, a future research direction is proposed which would involve finding optimal neighbors by combining the numeric distance and shape distance methods.

LSTM (Long Short Term Memory) network is one of the cyclic neural networks (RNNs). The algorithm was first published by Sepp Hochreiter and Jurgen Schmidhuber in Neural Computing [7]. It has better performance than ordinary RNN in processing and predicting time series related data. Based on the excellent performance of LSTM network in time series, Jiang et al. [8], taking the daily data of Shanghai Composite Index and Dow Jones Index as the research object, respectively uses RNNS and LSTM to build the model. And then they have found through experiments that for the neural network model compared with the RNN model, the LSTM model fits the data well. However, this model is not still not that well applicable to Dow Jones Index. in fact, in order to accurately predict Dow Jones Index, more hidden parameters should be added into the model, such as the influence of policy information, etc. However, this parameter is extremely difficult to quantify, so the forecast for Dow Jones Index so far there is still no better solution.

In order to make up for the deficiency of single forecasting model, a new forecasting model-hybrid model is proposed. In this type, decomposition methods, such as empirical mode decomposition (EMD) [9] and singular spectrum analysis (SSA) [10], are usually combined with ML and DL based on financial time series prediction models [11-13]. In recent works [14, 15], the hybrid models have been proved to perform better than the single baseline model because of its improved ability to capture time series patterns. As there is a nonlinear relationship between the predicted price of agricultural products and the influencing factors, Jia et al. [16] have designed a neural network model of LSTM-DA(Long Short-Term Memory-Double Attention) which combines the convolutional attention network, the Long Short-Term Memory network and the attention mechanism. Compared with the traditional signal model, this model can improve the prediction accuracy, and the predicted price index can accurately describe the overall trend of vegetable products in the next week.

However, hybrid models are rarely used to forecast DJIA prices. Meanwhile, a large number of studies have proved that denoising the high-frequency time series can effectively improve the generalization ability of the model and dramatically improve the prediction results. At present, the empirical data decomposition and noise reduction methods mainly include Ensemble Empirical Mode Decomposition (EEMD), Singular Spectrum Analysis (SSA) and Wavelet Transform Decomposition (WT). Although the integrated empirical mode decomposition (EEMD) can suppress the mode aliasing problem to some extent, the added white noise is neither completely neutralized nor entire. On the contrary, it may increase the complexity of the sequence. Jung et al. [17] have presented an integrated system where wavelet transforms and recurrent neural network (RNN) based on artificial bee colony (abc) algorithm (called ABC-RNN) are combined for stock price forecasting. Simulation results show that the proposed model with wavelet-based preprocessing greatly outperforms the other methods in TAIEX. However, it still has some insufficiencies. For example, it would be better to simplify the system organization, possibly by finding a scheme that includes functions of feature selection and an information supply of addictive parameters. In addition, a more advanced pattern selection scheme might be embedded in the system to retrieve significant patterns from the data.

This paper proposes an ensemble method based on data denoising methods, including the wavelet transform (WT) and singular spectrum analysis (SSA), and long-term short-term memory neural network (LSTM) to build a data prediction model . At the same time, we compare results of different models. The second part of this paper will introduce the model formula used in this paper, the third part will introduce the data, model prediction accuracy and stability results comparison and the fourth part is a summary.

## 2. Model Formulation

In this section, we provide an overview of the models used in this study, including the LSTM, WT, hybrid WT-LSTM, SSA and hybrid SSA-LSTM models.

### 2.1. Long term short-term memory

LSTM neural network was first proposed by Hochreiter and Schmidhuber , which is widely used to process sequence information owning to its advantages in discovering long-term dependencies. Therefore, it is theoretically feasible to establish an LSTM neural network model for financial high-frequency time series data. The structure of each neuron in LSTM is shown in Figure 1, and its internal structure includes a Cell and three Gates. Cell records the state of neurons, and Input Gate and Output Gate are used to receive, output and modify parameters. Forget Gate is used to control the forgotten degree of the previous unit state. The selection of activation function is an important part in the process of training neural network, which can make the neural network learn the nonlinear factors in the data. The activation function used in this paper is the commonly used tanh function.

In order to prove the effectiveness and versatility of the wavelet transform filtering method, the most common LSTM neural network structure is adopted in this research. The characteristic quantity is selected as the most basic highest price in the first five minutes, lowest price in the first five minutes, closing price in the first five minutes and opening price in the first five minutes. Specifically, the main structure of the LSTM neural network in this paper includes a 150-node LSTM layer, a 50-node LSTM layer, a fully connected layer and a dropout layer. The calculation diagram structure of LSTM neural network constructed in this paper is shown in Figure 2. The dotted box represents the neural network structure.

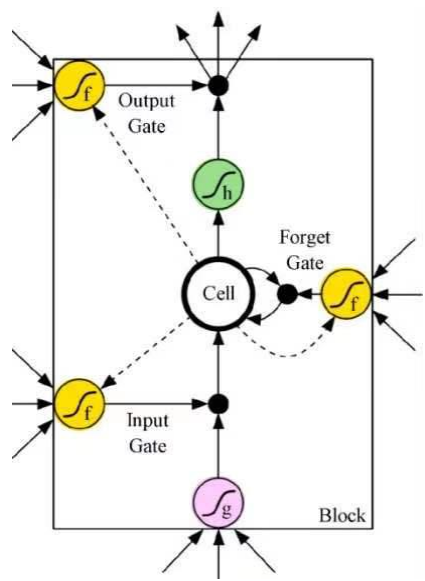

Fig. 1.  LSTM neuron structure

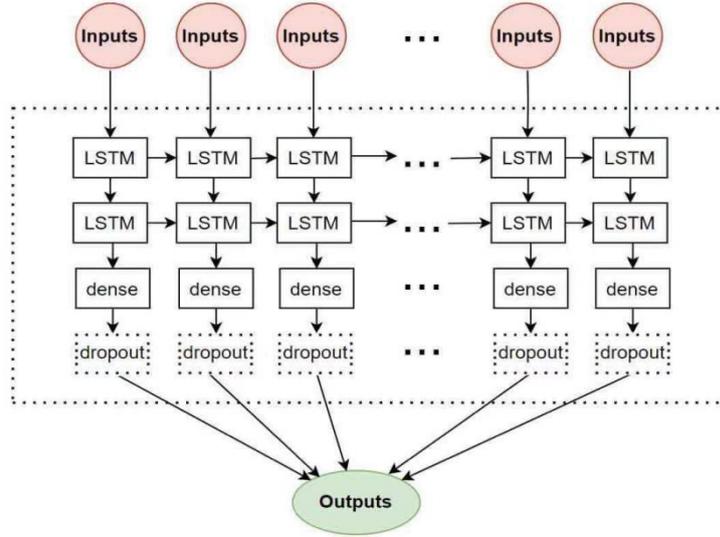

Fig. 2. LSTM deep neural network structure

## 2.2. Wavelet transform

Dow Jones Index is susceptible to a large number of factors such as politics, economy, and investor psychology. They usually contain a lot of noise and also show significant nonlinear characteristics. In order to improve the generalization ability of the model, the noise data should be filtered out when the deep neural network is used to process the nonlinear data. Wavelet analysis can carry out multi-scale refined analysis of signals through operation functions such as stretching and shifting, effectively eliminating noise in financial time series and fully retaining the characteristics of original signals [18]. Therefore, this paper proposes to use wavelet decomposition and reconstruction for data preprocessing of financial time series, and adopt "wavelet denoising" to eliminate the high-frequency components of noise in time series, so as to weaken the influence of short-term noise disturbance on neural network structure and improve the prediction ability of the model.

Wavelet decomposition decomposes each input signal into a low-frequency signal and a high-frequency signal, and only decomposes the low-frequency part. Assume that $C_0$ is the original financial time series signal, $C_1, C_2 \cdots C_l$ and $D_1, D_2 \cdots D_l$ is the first, second, and... , L-layer low frequency and high frequency signals. Then it can be mathematically expressed as:

$$C_0 = C_l + D_l + D_{l-1} + \cdots + D_2 + D_1 \tag{1}$$

In order to achieve the purpose of noise reduction, Mallat wavelet was reconstructed according to the low frequency coefficients of the first N layer and the high frequency coefficients of 1-N layer of wavelet decomposition, and the high frequency part of zero. The low frequency part of wavelet decomposition of financial time series data reflects the overall changing trend of the series, and the high frequency part reflects the short-term random disturbance of the series. Therefore, setting the high frequency part to zero can not only eliminate the noise and smooth the signal, but also obtain the approximate signal of the original financial time series data, so as to avoid the excessive learning of the neural network structure caused by short-term random disturbance factors and improve the extrapolation ability of the model.

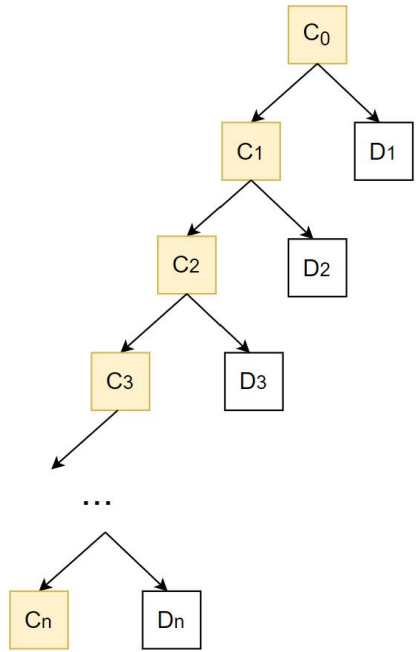

Fig. 3. Financial time series wavelet decomposition

*2.3. WT-LSTM*

In order to obtain higher accuracy, LSTM and WT were combined to predict the price of DJIA. Hybrid WT-LSTM consists of the following three phases and the modeling process is shown in the figure 4:

The first stage is WT decomposition: db wavelet or sym wavelet is often chosen as the wavelet basis for DJIA closing price forward fractional solution when dealing with economic, financial and temporal sequence data.

The second stage is time series reconstruction: to reconstruct the financial time series data and improve the generalization ability of the prediction model.

The third stage is LSTM Prediction: Smooted series $\tilde{x}_t$ and volume ($V_t$) are the input characteristics of LSTM. According to the partial autocorrelation function (PACF) of $x_t$, the time lag of $\tilde{x}_t$ and $V_t$ are determined. Then, the final prediction result is obtained.

The sym wavelet is an approximate symmetric orthogonal wavelet function of db wavelet, and it has better symmetry [19]. The more layers of wavelet decomposition, the better the stability of detail signal and approximate signal, however it will lead to greater errors in the decomposition process, so the number of layers should not be too much or too little. Therefore, in this paper, the sym4 wavelet basis is firstly used to divide the closing price of DJIA into four layers, so as to reconstruct the time series data of the gold melting and improve the generalization ability of the prediction model. On this basis, the general trend and market fluctuation information in the original time series are pre-processed. As a result, the hybrid WT-LSTM model proposed can avoid overfitting and obtain higher accuracy than the single LSTM model and the SSA-LSTM model.

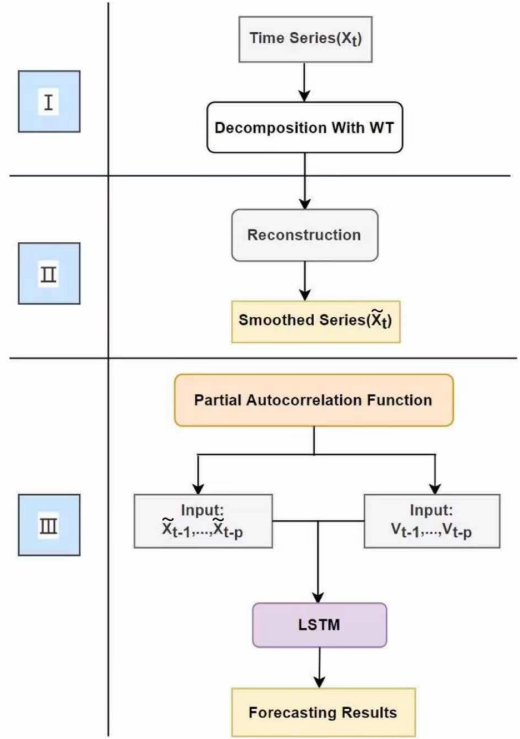

Fig. 4. hybrid WT-LSTM model processing process

## 2.4. Singular spectrum analysis and SSA-LSTM

In order to study the effects of the prediction accuracy of out-of-sample data and the prediction ability of future dynamic trend after the decomposition and reconstruction of financial time series data by wavelet transform (WT), we also carry out a controlled experiment to study the model prediction after financial time series data decomposition and reconstruction using singular spectrum analysis (SSA). The effect of SSA can construct a trajectory matrix from the observed financial time series, decompose and reconstruct the trajectory matrix, extract the different parts of the signal, thus effectively eliminating the noise of the financial time series and retaining the characteristics of the original signal.

Suppose there is a one-dimensional sequence $x(i)(i = 1,2,\cdots,n)$. Given that the embedding dimension is $m(m < \frac{n}{2})$, a time-delay matrix X can be obtained, and its dimension is $m \times k(k = n - m + 1)$.

$$X = \begin{bmatrix} x_1 & \cdots & x_{n-m+1} \\ \vdots & \ddots & \vdots \\ x_m & \cdots & x_n \end{bmatrix} \quad (2)$$

Let $S$ be the $m \times m$ dimensional covariance matrix of the delay matrix, then:

$$S = \begin{bmatrix} s(0) & \cdots & s(m-1) \\ \vdots & \ddots & \vdots \\ s(m-1) & \cdots & s(0) \end{bmatrix} \quad (3)$$

Singular spectrum analysis is used to decompose the covariance matrix $S$ to obtain $m$ singular values $\lambda_i$ ($i = 1,2,\cdots,m$). Then arrange the obtained $m$ singular values in descending order. The magnitude of the singular value represents the relative relationship between the signal and the noise. The singular value points with larger values are regarded as signal points, and the points with smaller values are regarded as noise points. The eigenvector $E^k$ corresponding to $\lambda_k$ is called the empirical orthogonal projection function. The orthogonal projection coefficient of the sampled signal $x(i)$ on the eigenvector $E^k$ is the $k$-th principal component:

$$a_i^k = \sum_{j=1}^{m} x_{i+j} E_j^k \ (0 \leqslant i \leqslant n - m) \tag{4}$$

If each principal component and the empirical orthogonal function are known, the process of reversing the original sequence is as follows:

$$x_{i+j} = \sum_{j=1}^{m} a_i^k E_j^k \ (1 \leqslant j \leqslant m) \tag{5}$$

So as to reach a comparison model with WT-LSTM model, we combined LSTM and SSA to predict the Dow price. The hybrid SSA-LSTM consists of the following three stages and the modeling process is shown in the figure 5:

The first stage is SSA Decomposition: SSA method is applied to decompose the original Bitcoin price time series ($x_t$) into three types of signals--trend, market volatility and noise.

The second stage is Time Series Reconstruction: reconstruction of the smooth sequence based on the trend and market fluctuation signal ($\tilde{x}_t$).

The third stage is LSTM Prediction: Smooted series $\tilde{x}_t$ and volume ($V_t$) are the input characteristics of LSTM. According to the partial autocorrelation function (PACF) of $x_t$, the time lag of $\tilde{x}_t$ and $V_t$ are determined. Then, the final prediction result is obtained.

SSA is used as a pre-treatment method to extract effective information about general trend and market fluctuation from original time series.

This paper chooses $m = 10$, that is, it is decomposed into ten layers. It is found that for this financial high-frequency time series, the first layer already contains more than 99.99% of the sequence information. Therefore, the first layer is selected as the reconstruction of the sequence, and the rest are used as noise.

Table 1. Singular values of singular spectrum analysis

| $i$ | $\lambda_i$ |
|---|---|
| 1 | 0.999991285112503 |
| 2 | 5.18930589813891e-06 |
| 3 | 1.42616024536684e-06 |
| 4 | 6.95916781349973e-07 |
| 5 | 4.21779000682764e-07 |
| 6 | 2.87333215462852e-07 |
| 7 | 2.15658198719760e-07 |
| 8 | 1.77801416057950e-07 |
| 9 | 1.55528979155002e-07 |
| 10 | 1.45403762506482e-07 |

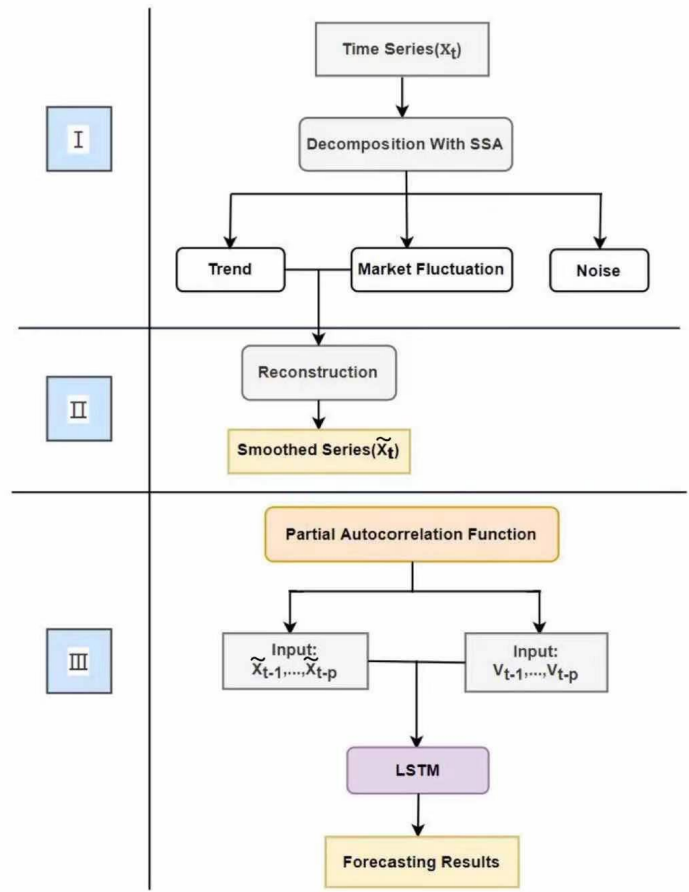

Fig. 5. hybrid SSA-LSTM model processing process

## 2.5. Training method and optimizer selection

The goal of this article is to compare the prediction effect of the closing price of Dow Jones Index, so mean square error (MSE) is selected as the loss function. As for the optimizer, this article uses the Adam optimizer (Adaptive Moment Estimation) for optimization training. Compared with other adaptive learning rate algorithms, the Adam algorithm has a faster convergence speed and a more effective learning effect. Epochs are set to be 10. This article is based on the Python language environment, and uses Tensorflow as the deep learning framework for training, prediction and comparison.

## 3. Data and Results
### 3.1. Data

In order to explore the applicability and effectiveness of the wavelet transform to predict the actual financial time series data, this section applies the financial forecasting model based on singular spectrum analysis and LSTM neural network to the DJIA closing price every 5 minutes to forecast high-frequency data of closing price and compare it with the result without reducing the noise. The data sample has selected the latest data available in the past years, that is, the time interval is from 00:00 on January 1, 2020 to 23:55 on December 31, 2020, with a time interval of 5 minutes, and a total of 19,666 pieces of data. The data comes from the wind database.

Table 2.  Descriptive statistics of DJIA closing price every 5 minutes

| Obs | Mean | Std.Dev | Variance | Skewness | Kurtosis |
|---|---|---|---|---|---|
| 19666 | 26876 | 2513.763 | 6319004 | -0.8633175 | 3.213287 |

### 3.2. Test set prediction effect evaluation index

To verify the validity of the model, the prediction and verification have been carried out from the two dimensions of short term (1 hour), medium term (3 hours) and long term (6 hours). Root mean square error (RMSE), mean absolute error (MAE) and mean absolute percentage error (MAPE) are used as the prediction accuracy indexes to evaluate the prediction effect of the test set. The smaller the values of the above three indicators are, the higher the prediction accuracy is. The prediction stability is evaluated by the standard deviation of absolute percentage error (SDAPE). The lower the SDAPE value, the higher the stability of prediction.

$$\text{RMSE} = \sqrt{\frac{1}{N}\sum_{i=1}^{N}(y_i - \widehat{y}_i)^2} \quad (5)$$

$$\text{MAE} = \frac{1}{N}\sum_{i=1}^{N}|y_i - \widehat{y}_i| \quad (6)$$

$$\text{MAPE} = \frac{1}{N}\sum_{i=1}^{N}|\frac{y_i - \widehat{y}_i}{y_i}| \times 100 \quad (7)$$

$$\text{SDAPE} = \sqrt{\frac{1}{N}\sum_{i=1}^{N}(\left|\frac{y_i - \widehat{y}_i}{y_i}\right| - MAPE)^2} \quad (8)$$

Here, $y_i$ is the actual value and $\widehat{y}_i$ is prediction obtained from the predicting model. $N$ is the number of prediction.

### 3.3. Comparative analysis of short-term forecasting effects

Table 3.  1-hour DJIA closing price forecast results

|   | RMSE | MAE | MAPE | SDAPE |
|---|---|---|---|---|
| LSTM | 5.8516916 | 4.5195833 | 0.0001481 | 0.0001218 |
| Dropout-LSTM | 3.8146496 | 2.8042500 | 0.0000919 | 0.0000848 |
| SSA-LSTM | 1.7488158 | 1.5490332 | 0.0000508 | 0.0000266 |
| WT-LSTM | 1.1966503 | 1.0434276 | 0.0000342 | 0.0000192 |

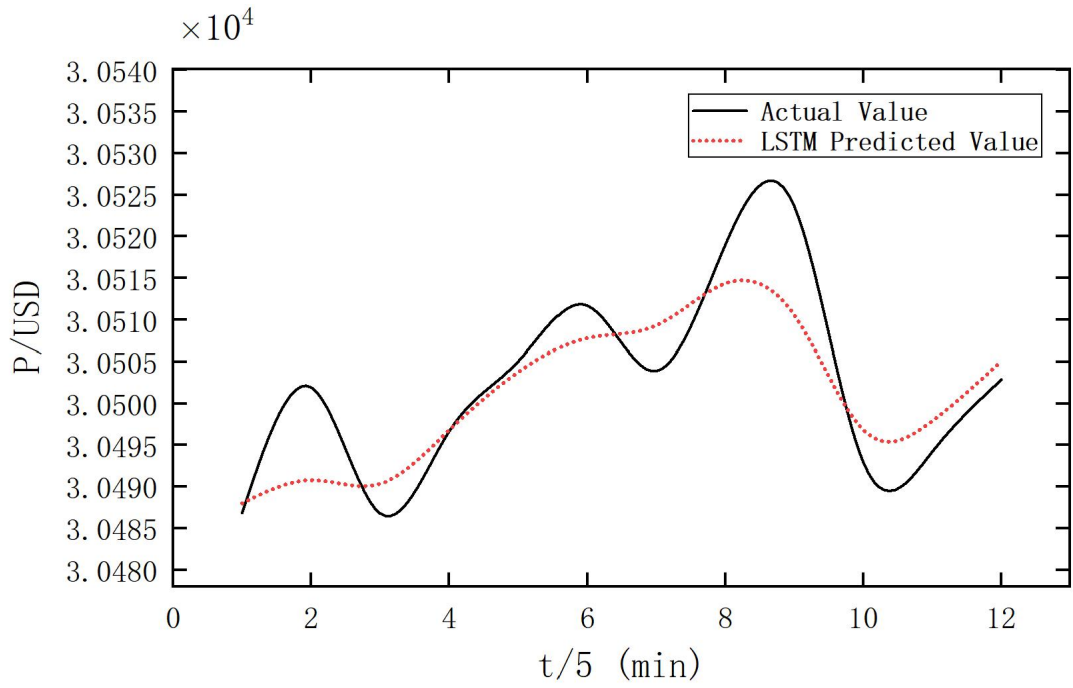

Fig. 6.  Short-term LSTM prediction

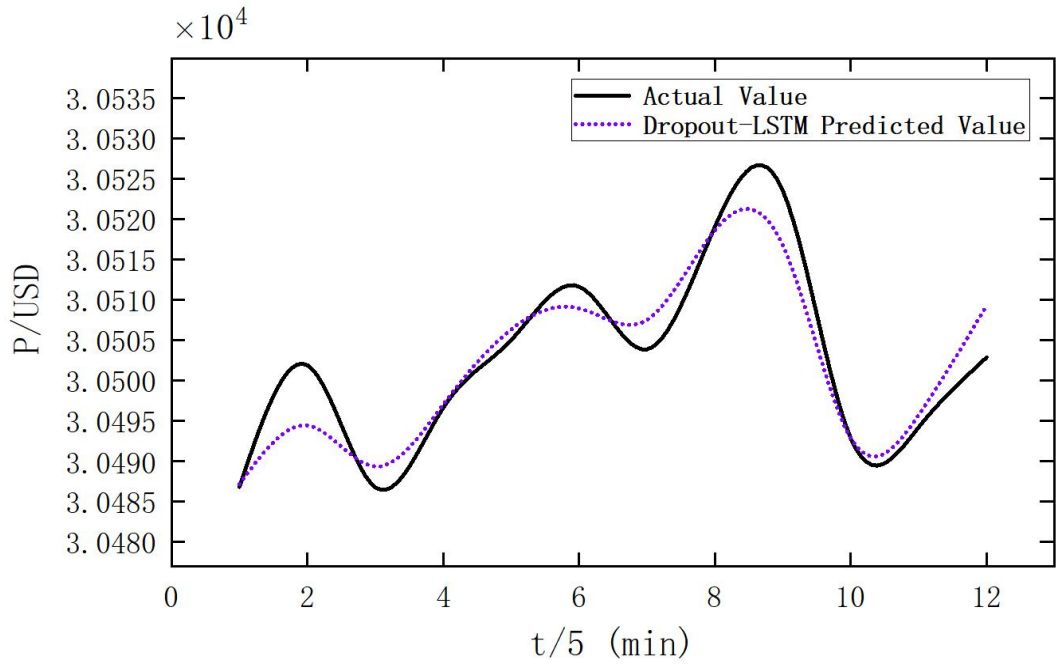

Fig. 7. Short-term Dropout-LSTM prediction

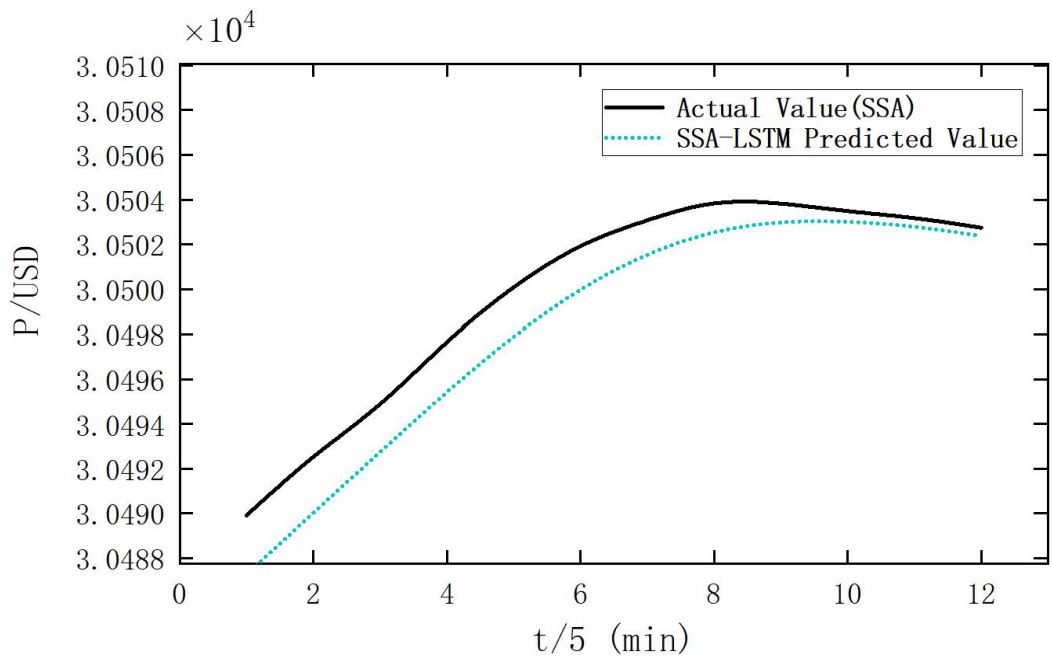

Fig. 8. Short-term SSA-LSTM prediction

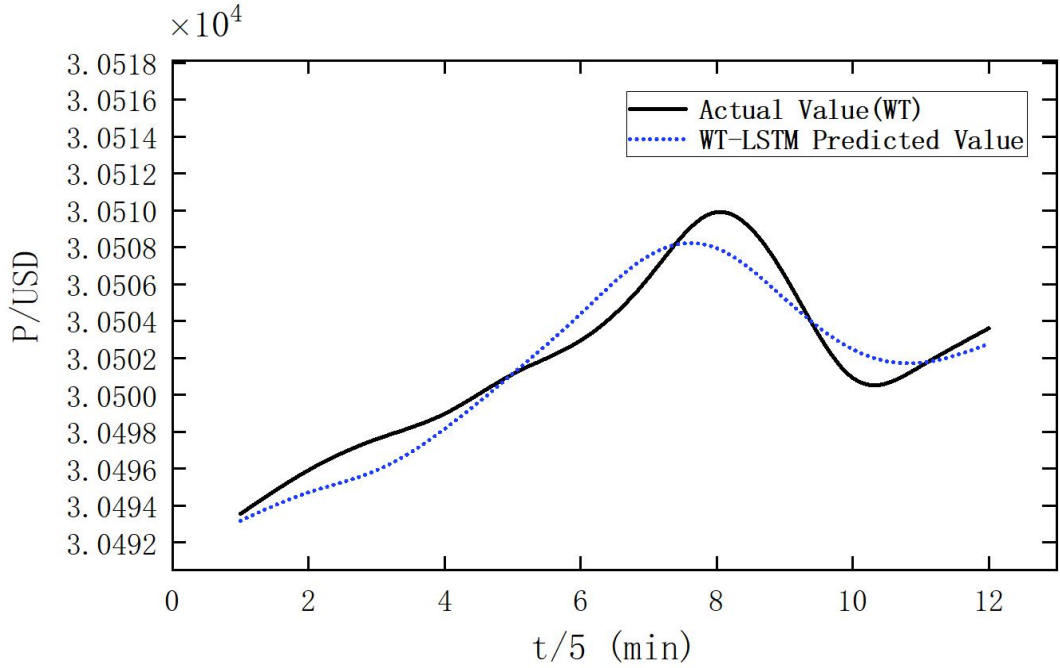

Fig. 9. Short-term WT-LSTM prediction

For the short-term prediction effect, in terms of prediction accuracy, Dropout [20] can improve the prediction effect of the original LSTM model, and RMSE, MAE and MAPE are reduced by 34.81%, 37.95% and 37.95%, respectively. SSA-LSTM can improve the prediction effect of the original LSTM model, RMSE, MAE and MAPE decreased by 70.11%, 65.73% and 65.72%, respectively. WT-LSTM can improve the prediction effect of the original LSTM model, and RMSE, MAE and MAPE can be reduced by 79.55%, 76.91% and 76.91%, respectively. With regard to predictive stability, Dropout can improve the predictive stability and reduce the SDAPE by 30.41%, and SSA-LSTM can improve the predictive stability and reduce the SDAPE by 78.14%. WT-LSTM can improve the stability of the original prediction and reduce the SDAPE by 84.23%. Through the analysis of short-term prediction, it is found that the generalization effect of filtering to prevent overfitting is better than the effect of Dropout to improve the accuracy. At the same time, the wavelet transform has a good effect in filtering. In terms of prediction accuracy, WT-LSTM can improve the prediction effect of SSA-LSTM model, and reduce RMSE, MAE and MAPE by 31.57%, 32.64% and 32.64%, respectively. In respect of predictive stability, WT-LSTM can improve the predictive stability of SSA-LSTM model and reduce SDAPE by 27.85%. The prediction results of the four methods are shown in Figure 6-9.

### *3.4. Comparative analysis of medium-term forecasting effects*

Table 4. 3-hour DJIA closing price forecast results

|  | RMSE | MAE | MAPE | SDAPE |
| --- | --- | --- | --- | --- |
| LSTM | 5.2447743 | 4.1066389 | 0.0001347 | 0.0001069 |
| Dropout-LSTM | 3.4334542 | 2.6221944 | 0.0000860 | 0.0000727 |
| SSA-LSTM | 1.1713269 | 0.9653596 | 0.0000317 | 0.0000222 |
| WT-LSTM | 1.2796409 | 1.0970283 | 0.0000360 | 0.0000216 |

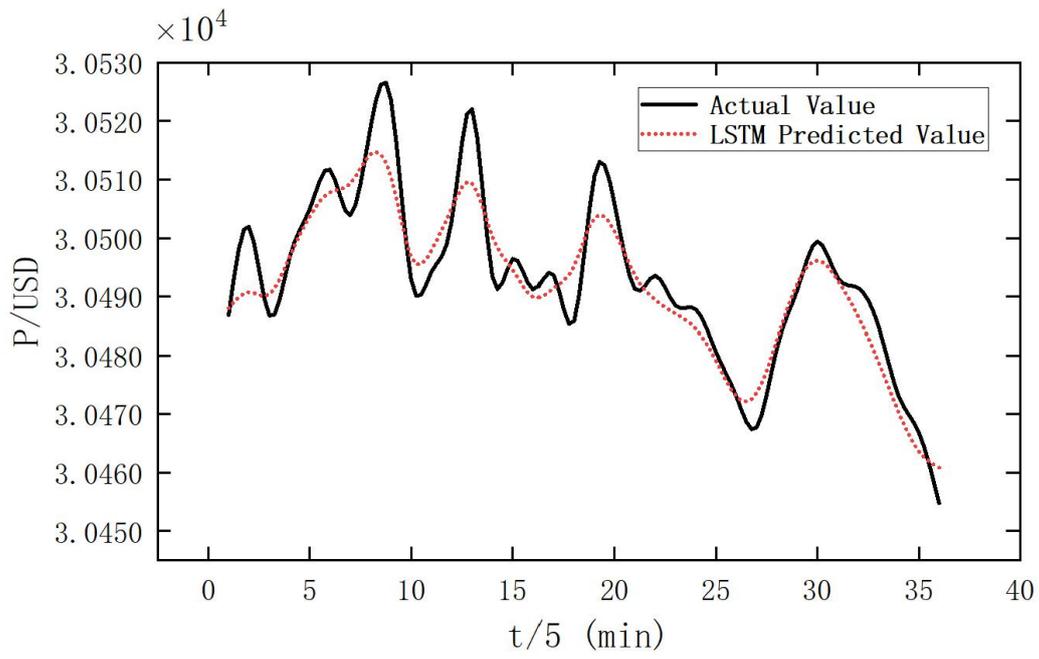

Fig. 10. Medium-term LSTM prediction

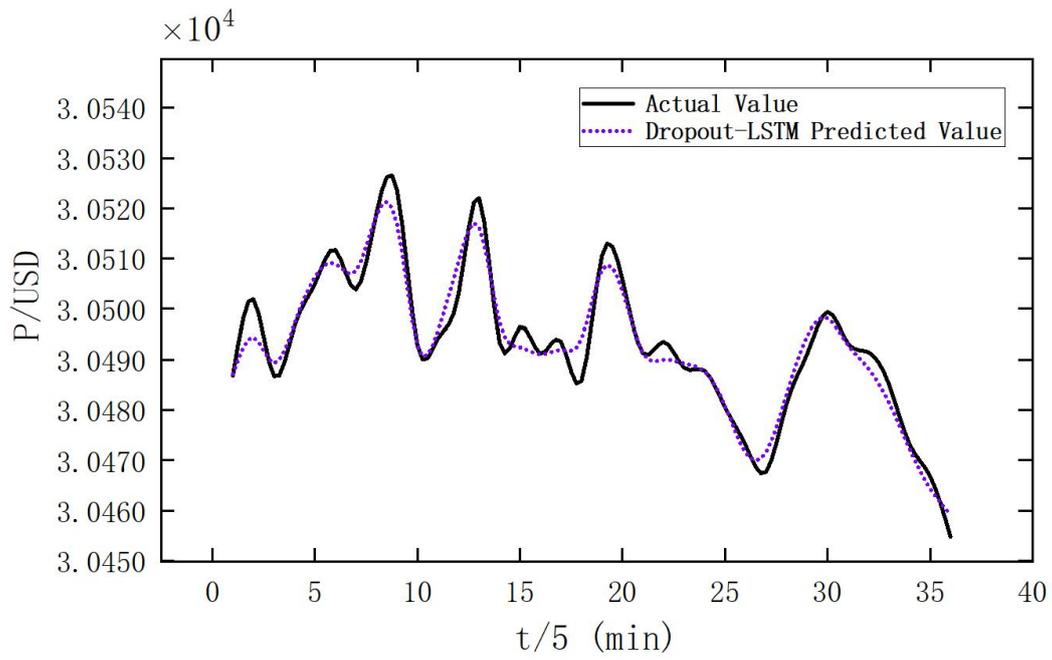

Fig. 11. Medium-term Dropout-LSTM prediction

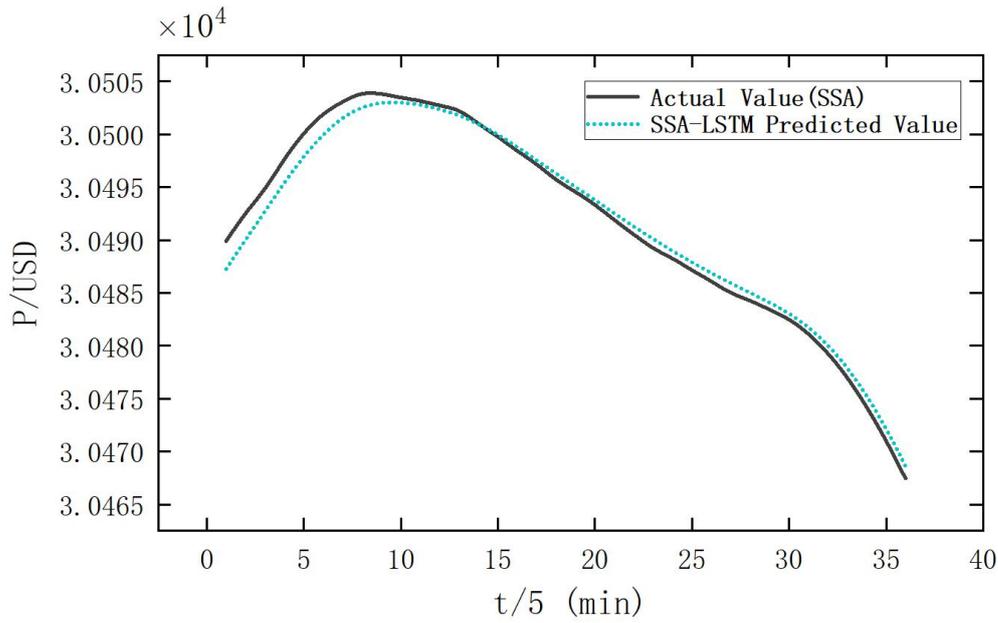

Fig. 12. Medium-term SSA-LSTM prediction

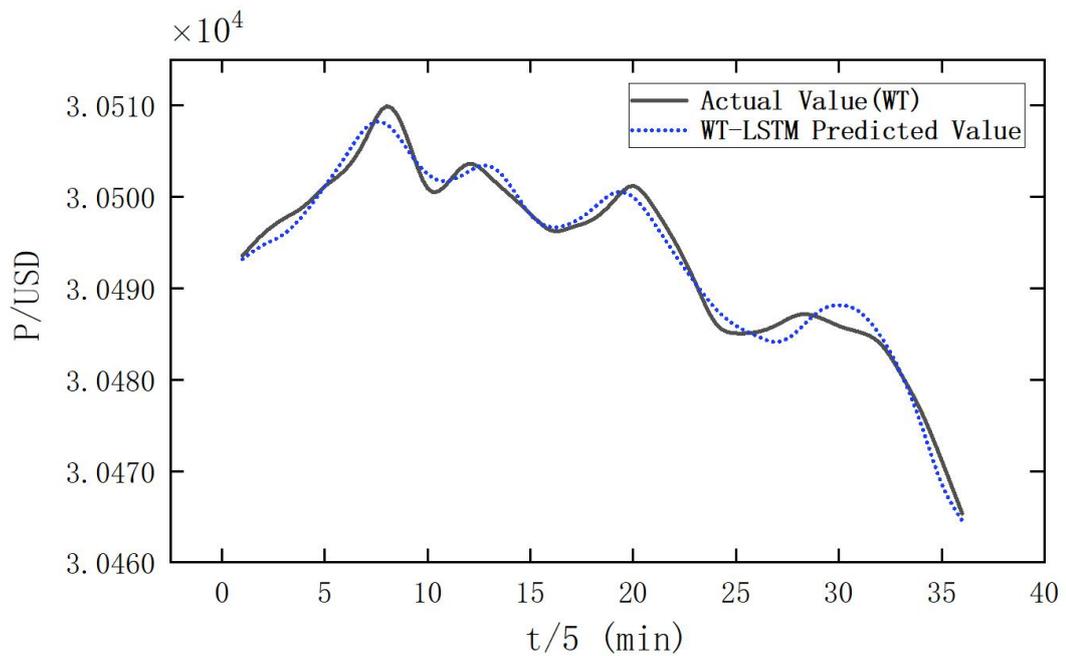

Fig. 13. Medium-term SSA-LSTM prediction

For the medium-term prediction effect, in terms of prediction accuracy, Dropout can improve the prediction effect of the original LSTM model, and RMSE, MAE and MAPE are reduced by 34.54%, 36.15% and 36.14%, respectively. SSA-LSTM can improve the prediction effect of the original LSTM model, RMSE, MAE and MAPE decreased by 77.67%, 76.49% and 76.49%, respectively. WT-LSTM can improve the prediction effect of the original LSTM model, and reduce RMSE, MAE and MAPE by 75.60%, 73.29% and 73.28%, respectively. In respect of predictive stability, Dropout can improve the predictive stability and reduce the SDAPE by 32.03%, and SSA-LSTM can improve the predictive stability and reduce the SDAPE by 79.20%. WT-LSTM can improve the stability of the original prediction and reduce the SDAPE by 79.79%. Through the analysis of short-term prediction, we can see that the generalization effect of filtering to prevent overfitting is better than the effect of Dropout to improve the accuracy. At the same time, Singular Spectrum Analysis also has a good effect in filtering. In terms of prediction accuracy, SSA-LSTM can improve the prediction effect of WT-LSTM model, and reduce RMSE, MAE and MAPE by 8.46%, 12.00% and 12.00%, respectively. With regard to predictive stability, WT-LSTM can improve the predictive stability of SSA-LSTM model and reduce SDAPE by 2.84%. The predicted results of the four methods are shown in Figure 10-13.

### *3.5. Comparative analysis of long-term forecasting effects*

Table 5.  6-hour DJIA closing price forecast results

|  | RMSE | MAE | MAPE | SDAPE |
| --- | --- | --- | --- | --- |
| LSTM | 6.1655946 | 4.5780000 | 0.0001503 | 0.0001356 |
| Dropout-LSTM | 4.5014469 | 3.2249583 | 0.0001059 | 0.0001032 |
| SSA-LSTM | 1.1753464 | 0.9942363 | 0.0000326 | 0.0000206 |
| WT-LSTM | 1.9164739 | 1.4123594 | 0.0000464 | 0.0000426 |

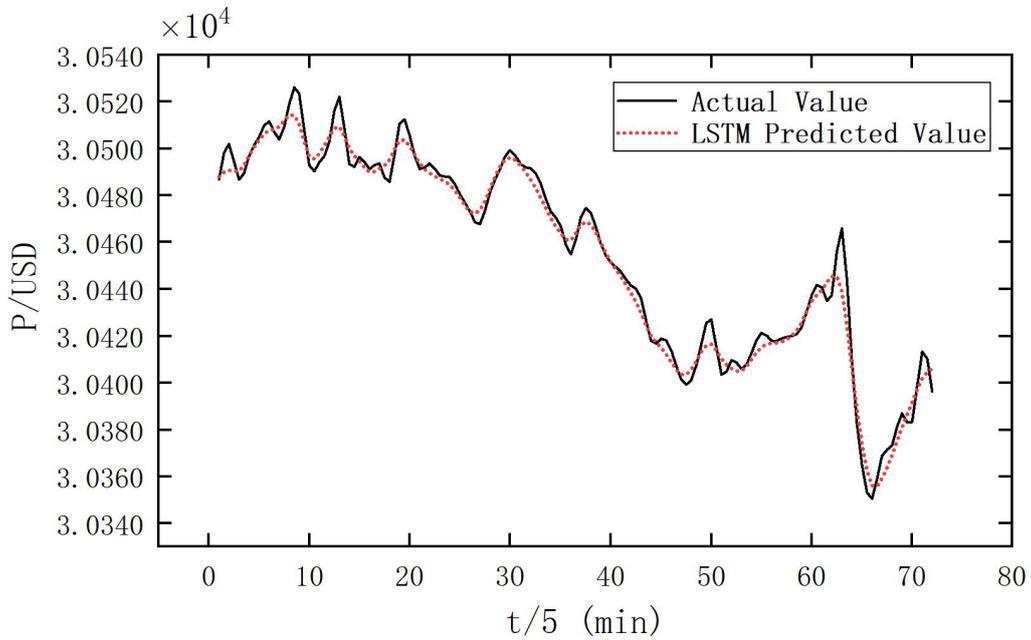

Fig. 14.  Long-term LSTM prediction

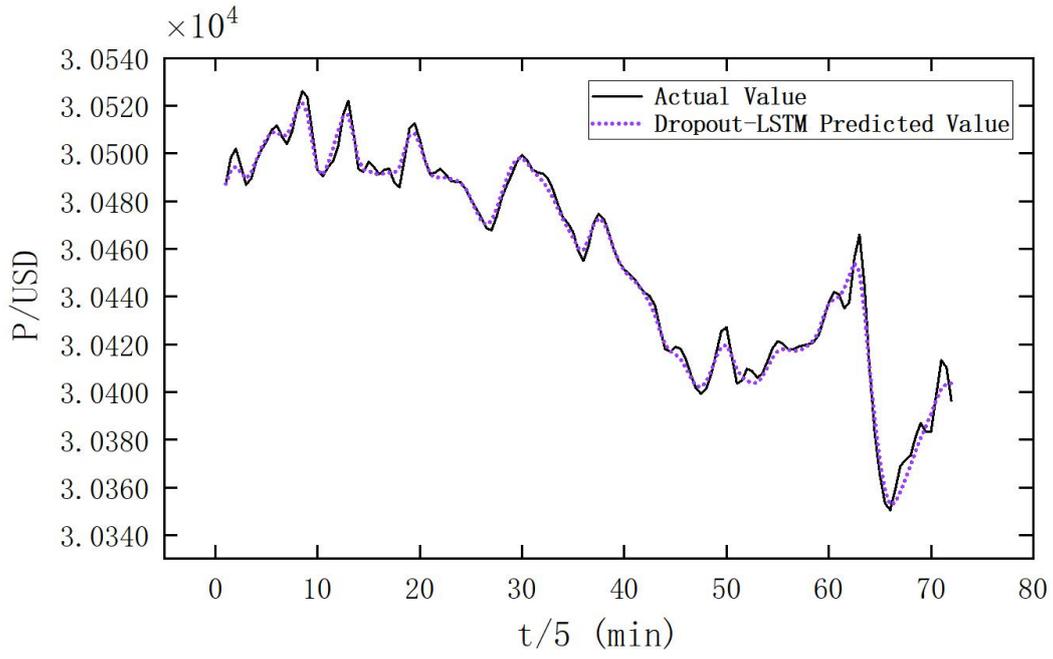

Fig. 15. Long-term Dropout-LSTM prediction

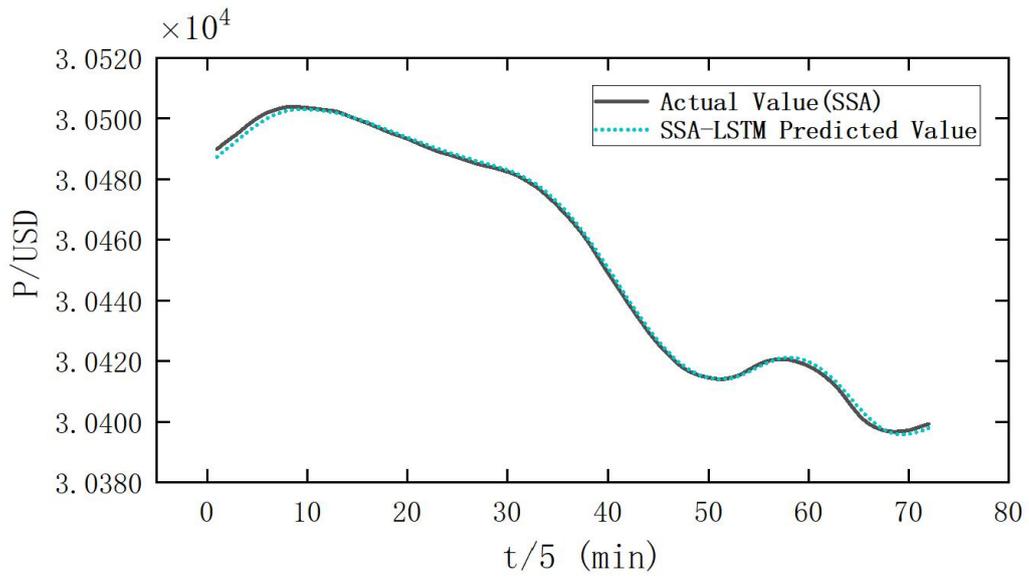

Fig. 16. Long-term SSA-LSTM prediction

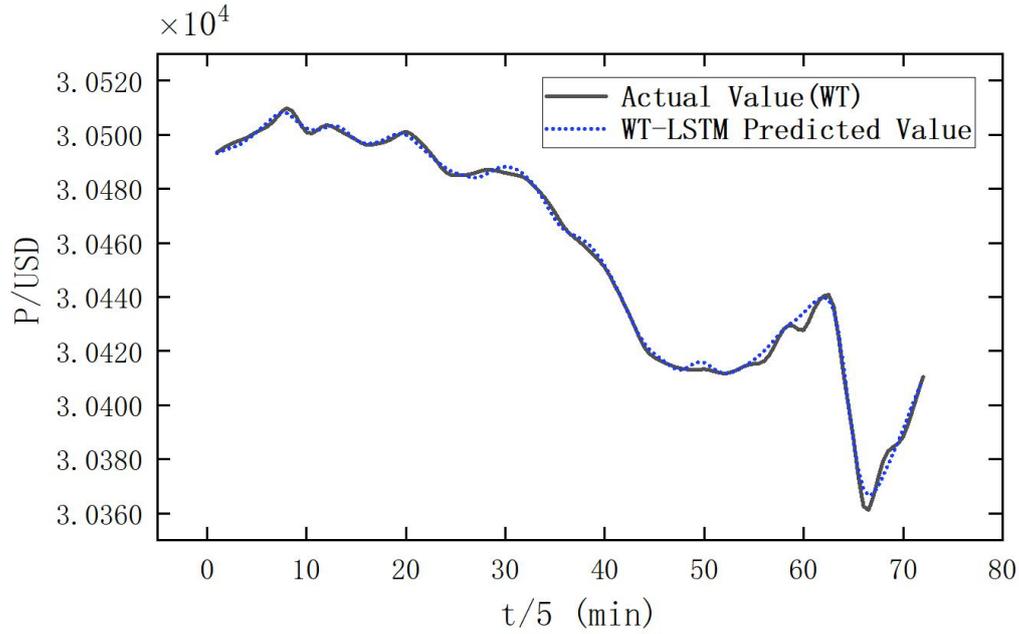

Fig. 17.  Long-term WT-LSTM prediction

For long-term forecasts, in terms of prediction accuracy, Dropout can improve the prediction effect of the original LSTM model, and RMSE, MAE and MAPE are reduced by 26.99%, 29.56% and 29.54%, respectively. SSA-LSTM can improve the prediction effect of the original LSTM model, RMSE, MAE and MAPE decreased by 80.94%, 78.28% and 78.28%, respectively. WT-LSTM can improve the prediction effect of the original LSTM model, and reduce RMSE, MAE and MAPE by 68.92%, 69.15% and 69.14%, respectively. In terms of predictive stability, Dropout can improve the predictive stability and reduce the SDAPE by 23.90%, and SSA-LSTM can improve the predictive stability and reduce the SDAPE by 84.82%. WT-LSTM can improve the stability of the original prediction and reduce the SDAPE by 68.57%. According to the analysis of short-term prediction, we can see that the generalization effect of filtering to prevent overfitting is better than the effect of Dropout to improve the accuracy. At the same time, Singular Spectrum Analysis also has a good effect in filtering. In terms of prediction accuracy, SSA-LSTM can improve the prediction effect of WT-LSTM model, and reduce RMSE, MAE and MAPE by 38.67%, 29.60% and 29.63%, respectively. In terms of predictive stability, SSA-LSTM can improve the predictive stability of WT-LSTM model and reduce SDAPE by 51.70%. The predicted results of the four methods are shown in Figure 14-17.

In summary, both WT-LSTM and SSA-LSTM can significantly improve the prediction ability of the original LSTM, and improve the prediction accuracy, stability and generalization ability, so as to accurately predict the changing trend of the Dow price, no matter in the short, medium or long term. Also, in the short term and medium term, the improvement effect of WT-LSTM is better than that of SSA-LSTM, while in the long term, the improvement effect of SSA-LSTM is better than that of WT-LSTM. The comprehensive comparison of the prediction results of the four methods is shown in Figure 18-20.

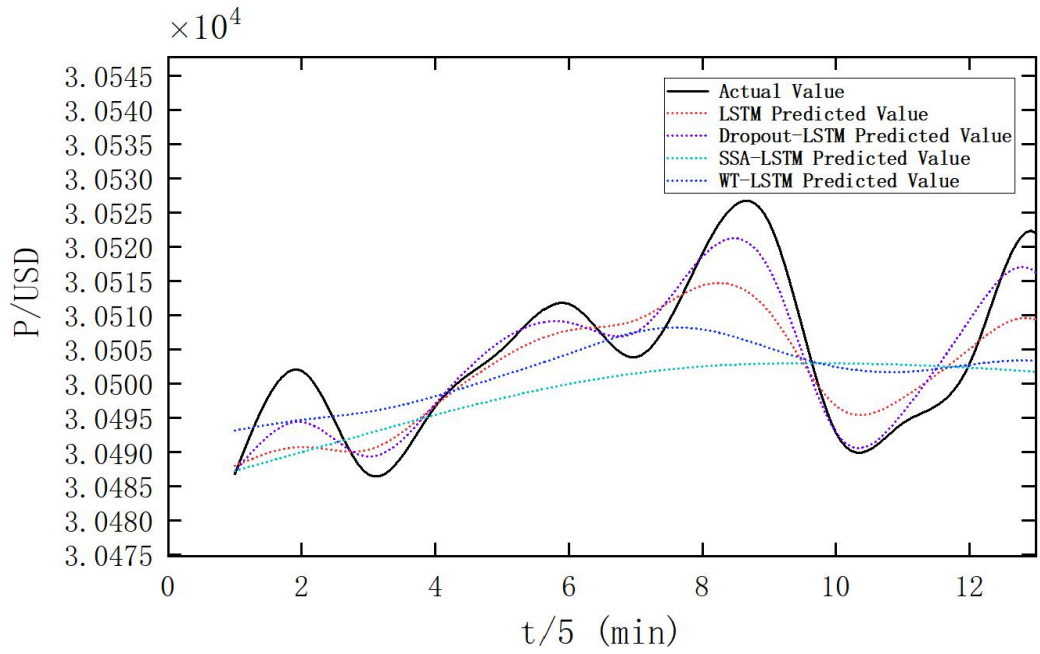

Fig. 18. Short-term prediction results

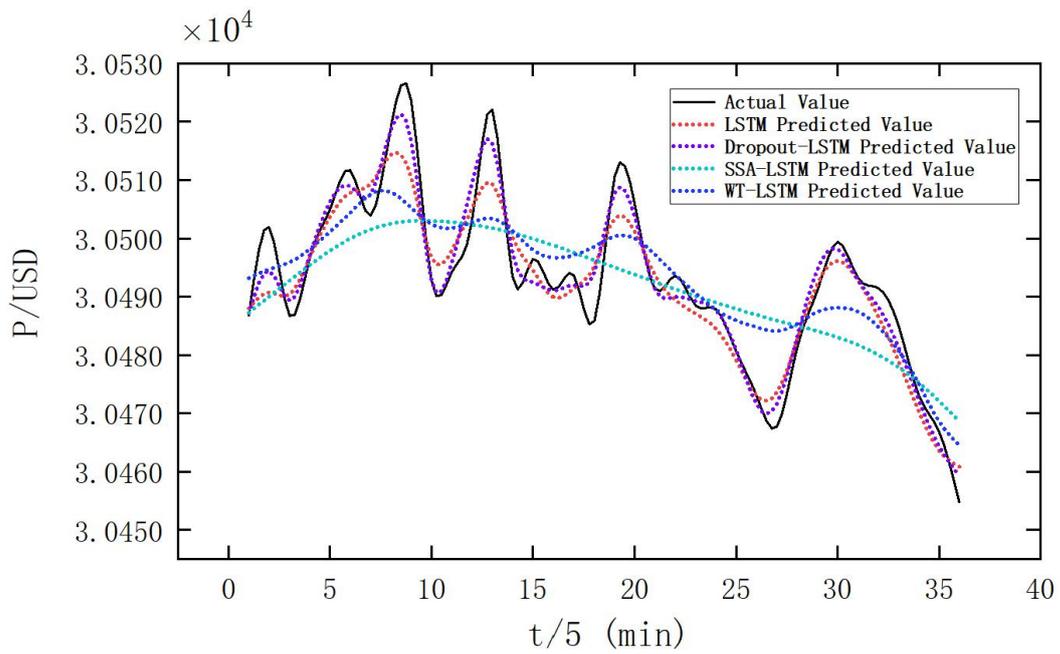

Fig. 19. Medium-term prediction results

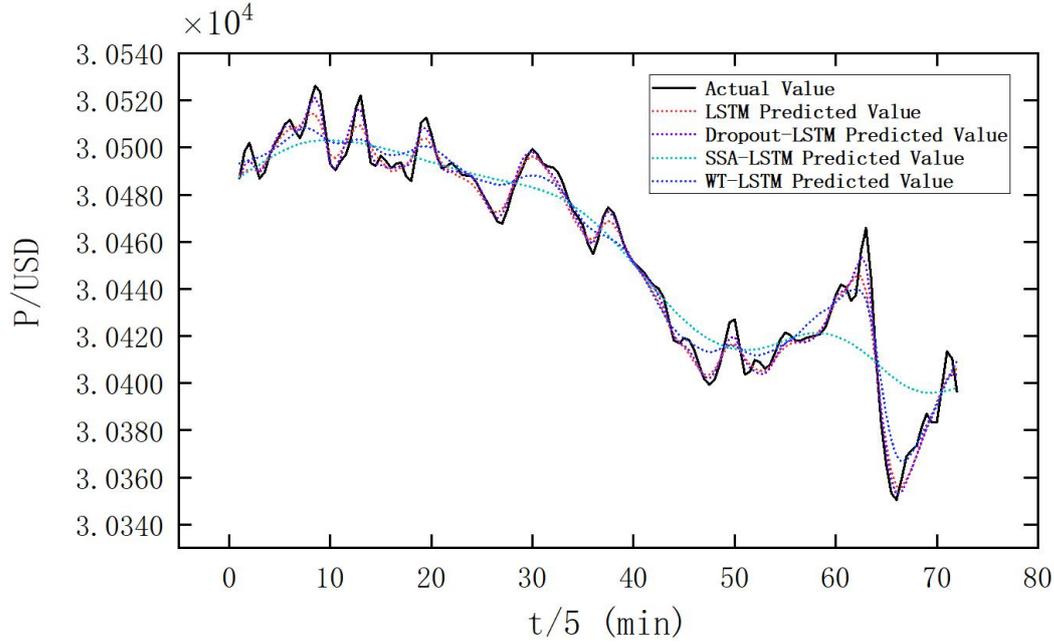

Fig. 20. Long-term prediction results

**4. conclusion and discussions**

This paper has discussed the theoretical basis of deep learning and the practical application of LSTM price prediction neural network, and has proposed the use of denoising methods to reduce noise on high frequency financial time series to reduce the impact of random interference noise to improve the prediction accuracy of the model for out-of-sample data and the ability to predict future dynamics. The results are also significant enough to prove the improvement of LSTM predicting model with effective denoising methods, including wavelet transform and singular spectrum analysis.

    Based on the empirical results of the Dow 5 minutes closing data, the following conclusions can be drawn: firstly, the use of wavelet transform and singular spectrum analysis to denoise data can significantly improve the generalization ability of LSTM neural network, and WT's effect is better than SSA's effect in the short term and medium term, but worse than SSA filtering method in the long term. Secondly, with the extension of time limit, the generalization ability of wavelet transform and reconstructed filter sequence in the prediction of LSTM neural network is weakening, while the generalization ability of singular spectrum analysis decomposition and reconstructed filter sequence in the prediction of LSTM neural network is increasing. But the prediction effect of wavelet transform and singular spectrum analysis reconstruction filter is still significant. Thirdly, the WT-LSTM neural network and SSA-LSTM neural network can converge quickly in a small amount of time, and has a good prediction effect under the high frequency data, which provides a new idea for the financial risk management and monitoring under the high frequency trading.

    In view of the high tunability of neural network, there are still many technical improvements in the future research, such as adding more non-homogeneous information as input to the neural network and optimizing the structure of the neural network itself. It is worth noting that applying the advantages of large data in financial high-frequency time series to the investment field can help investors quickly identify investment opportunities and promote the development of intelligent investment in financial markets. In addition, it can also strengthen risk management, improve the efficiency of risk

identification, and effectively maintain the stability of the financial market.

**Acknowledgements**

We would like to thank Southeast University for providing us with a strong academic atmosphere and rich academic resources, which enables us to complete our academic papers more efficiently. At the same time, I would like to thank the partners for their full cooperation and joint discussion, which made our professional ability and knowledge to a new level in this kind of academic research.